\documentclass[twocolumn,english,superscriptaddress,aps,prb]{revtex4-1}
\usepackage[T1]{fontenc}
\usepackage[latin9]{inputenc}
\usepackage{verbatim}
\usepackage{amsmath}
\usepackage{amssymb}
\usepackage{graphicx}
\usepackage{esint}
\usepackage{mathtools}

\makeatletter
\@ifundefined{textcolor}{}
{%
 \definecolor{BLACK}{gray}{0}
 \definecolor{WHITE}{gray}{1}
 \definecolor{RED}{rgb}{1,0,0}
 \definecolor{GREEN}{rgb}{0,1,0}
 \definecolor{BLUE}{rgb}{0,0,1}
 \definecolor{CYAN}{cmyk}{1,0,0,0}
 \definecolor{MAGENTA}{cmyk}{0,1,0,0}
 \definecolor{YELLOW}{cmyk}{0,0,1,0}
 }

\usepackage[T1]{fontenc}
\usepackage[latin9]{inputenc}
\usepackage{verbatim}
\usepackage{amsmath}
\usepackage{amssymb}
\usepackage{esint}
\usepackage{graphicx} %%%sid

\makeatletter
\@ifundefined{textcolor}{}
{%
 \definecolor{BLACK}{gray}{0}
 \definecolor{WHITE}{gray}{1}
 \definecolor{RED}{rgb}{1,0,0}
 \definecolor{GREEN}{rgb}{0,1,0}
 \definecolor{BLUE}{rgb}{0,0,1}
 \definecolor{CYAN}{cmyk}{1,0,0,0}
 \definecolor{MAGENTA}{cmyk}{0,1,0,0}
 \definecolor{YELLOW}{cmyk}{0,0,1,0}
 }

\makeatother

\newcommand{\beq}{\begin{equation}}
\newcommand{\eeq}{\end{equation}}
\newcommand{\bea}{\begin{eqnarray}}
\newcommand{\eea}{\end{eqnarray}}

\newcommand{\homega}{\hat{\omega}}
\newcommand{\hepsilon}{\hat{\epsilon}}
\newcommand{\omegan}{\omega_n}

\def\zhat{{\hat z}}

\def\br{\mathbf{r}}
\def\bk{\boldsymbol{k}}
\def\bq{\boldsymbol{q}}

\def\be{\begin{eqnarray}}
\def\ee{\end{eqnarray}}

\def\inv{\mathcal{I}}
\def\tr{\mathcal{T}}

\def\urlprefix{}
\def\url#1{}

\usepackage{babel}

\makeatother

\usepackage{babel}
\begin{document}

\title{Charge Transport in Weyl Semimetals}

\author{Pavan Hosur}

\affiliation{Department of Physics, University of California at Berkeley, Berkeley,
CA 94720}

\author{S. A. Parameswaran}

\affiliation{Department of Physics, University of California at Berkeley, Berkeley,
CA 94720}

\author{Ashvin Vishwanath}

\affiliation{Department of Physics, University of California at Berkeley, Berkeley,
CA 94720}

\affiliation{Materials Science Division, Lawrence Berkeley National Laboratories,
Berkeley, CA 94720}

\begin{abstract}
We study transport  in three dimensional Weyl semimetals with $N$ isotropic Weyl nodes in the
presence of Coulomb interactions or disorder at temperature $T$.
In the interacting clean limit, we determine the conductivity by solving a quantum Boltzmann equation within a `leading log' approximation and find it to be proportional to $T$, upto logarithmic factors arising from the flow of couplings. In the noninteracting disordered case, we compute the finite-frequency Kubo conductivity and show that it  exhibits distinct behaviors for  $\omega\ll T$  and $\omega\gg T$: in the former regime we recover the results of a  previous analysis, of a finite conductivity and a Drude width that vanishes as $NT^2$; in the latter, we find a conductivity that vanishes linearly with $\omega$ whose leading contribution as $T\rightarrow 0$ is the same as that of the clean, non-interacting system $\sigma(\omega, T=0) = N\frac{e^2}{12h} \frac{|\omega|}{v_F}$. 
We compare our results to experimental data on Y$_2$Ir$_2$O$_7$ and also comment on the possible relevance to recent transport data on Eu$_2$Ir$_2$O$_7$.

\end{abstract}
\maketitle

There has been a surge of recent activity studying Dirac excitations in two dimensional media, most famously graphene \cite{GeimGraphene}. A natural question is whether there are analogs in three dimensions, with a vanishing density of states at the chemical potential and linearly dispersing excitations. It has long been known that touchings between a pair of non-degenerate bands are stable in three dimensions, and typically have linear dispersion. Near these, electronic excitations are described by an analog of the Weyl equation of particle physics, which describes two-component chiral fermions \cite{Herring, Abrikosov, NielsenABJ}. Hence these states have been dubbed {\em Weyl semi-metals} (WSMs) \cite{PyrochloreWeyl}.

To remove a band touching (or Weyl node) one necessarily must connect to another node. This is in contrast with two dimensions: graphene's nodes can be gapped by different intranode perturbations that break inversion ($\inv$) or time reversal  ($\tr$) symmetry. The enhanced protection in three dimensions is due to a topological property of the nodes - they are sources (monopoles) of Chern flux in the Brillouin zone (BZ).
This momentum space topology is associated with several physical phenomena. In particular, it was recently realized \cite{PyrochloreWeyl}  that unusual surface states will result as a consequence of the band topology. These take the form of Fermi arcs that connect the projections of the nodes onto the surface BZ. Such topological properties are sharply defined as long as one can distinguish band touching associated with opposite Chern flux. The presence of translation symmetry, and hence conserved crystal momenta, is sufficient to protect these defining properties since the nodes are separated  in the BZ. In principle one needs perfect crystalline order to define these phases; in practice, smooth disorder that only weakly mixes nodes is expected to have little effect. Other manifestations of the band topology include an anomalous Hall effect  \cite{RanQHWeyl, WeylMultiLayer} that is tied to the momentum space displacement between nodes, and magneto-resistance arising from Adler-Bell-Jackiw anomaly of Weyl fermions \cite{AjiABJAnomaly,
NielsenABJ}.

Physical realizations of WSMs require non-degenerate bands to touch; therefore spin degeneracy must be lifted (by either spin-orbit interactions or magnetic order), and either $\tr$ or $\inv$ must be broken: otherwise, all bands would be doubly degenerate.
We further require that the Fermi `surface' consists exactly of the Weyl nodes. In $\tr$-breaking realizations where $\inv$ is unbroken, a simple `parity criterion' applied to eight $\tr$-invariant momenta in the BZ can be used to diagnose the existence of Weyl nodes  \cite{TurnerAV}. In   \cite{PyrochloreWeyl}, certain  pyrochlore iridates A$_2$Ir$_2$O$_7$ (A$=$Y or Eu), were proposed to be magnetically ordered WSMs, with $N=24$ Weyl points, all at the Fermi energy;  \cite{Kim} reached similar conclusions but with $N=8$. Alternate proposals include HgCr$_2$Se$_4$ in the ferromagnetic state \cite{FangChernSemiMetal} and  topological insulator-ferromagnet heterostructures  \cite{WeylMultiLayer}, with $N=2$, the minimum  allowed.

Motivated by the availability of transport data on the iridates \cite{WeylResistivityMaeno,EuIridateExperiments},  we study the electrical conductivity of an idealized model of a WSM, with an even number $N$ of isotropic Weyl nodes characterized by the same dispersion, with $N/2$ nodes of each chirality as required by topology \cite{NielsenNinomiya,PyrochloreWeyl}.  In the absence of impurities and interactions we expect the free fermion result, $\sigma_0^{(N)}(\omega) =  N\frac{e^2}{12h} \frac{|\omega|}{v_F}$.  We demonstrate how this is modified in two cases:

\noindent(i) in clean undoped systems with Coulomb interactions, current is carried equally by counterpropagating electrons and holes and can be relaxed via interactions alone. Solving a quantum Boltzmann equation (QBE) we find a finite conductivity proportional to the temperature $T$ (upto logarithmic factors), as expected of a quantum critical system  \cite{FritzGrapheneConductivity}, where $T$ is the sole energy scale, 
\begin{equation}\label{eq:sigmadcint}
\sigma^{(N)}_{\rm dc}(T)= \frac{e^2}{h}\frac{k_BT}{\hbar v_F(T)} \frac{1.8}{\alpha_T^2\log\alpha_T^{-1}}
\end{equation}
Here $v_F(T)= v_F \left({\alpha_0}/{\alpha_T}\right)^{\frac{2}{N+2}}$ and $\alpha_T = {\alpha_{0}}\left[{1+\frac{(N+2)\alpha_{0}}{3\pi}\mbox{ln}\left(\frac{\hbar\Lambda}{k_{B}T}\right)}\right]^{-1}$
 are the  Fermi velocity and  fine structure constant renormalized to the scale of the temperature  $T$,
and $v_F$ and $\alpha_0=e^2/\varepsilon \hbar v_F$ are the corresponding `bare' values at the microscopic scale (See App. \ref{app:RG}). %$\Lambda$

\noindent(ii) In the presence of impurities, power counting shows that white-noise disorder is an irrelevant perturbation, and a naive expectation is that the clean result $\sigma^{(N)}_0$ is reproduced. However, the result is more interesting: by evaluating a standard Kubo formula,  we find that the the finite-frequency conductivity  exhibits different behaviors for  $\omega\ll T$  and $\omega\gg T$: in the former regime we find in agreement with \cite{WeylMultiLayer} a finite Drude-like response with a peak width vanishing as $N T^2$;  in the latter,  we recover  $\sigma^{(N)}_0$ as the leading behavior, which is universal and independent of disorder.  We also determine the manner in which the conductivity interpolates between these limits.

Previous studies of 3D Dirac points have assumed Lorentz invariance  \cite{LeadingLogTransport} or worked at a topological phase transition between insulators \cite{GoswamiDiracTransport}. Although our work differs from both of these situations -- instantaneous Coulomb interactions break Lorentz invariance, and we study a stable phase -- there are sufficient parallels that a similar `leading log' approximation suffices to solve the QBE. Coulomb interactions also lead to a finite dc conductivity in clean graphene -- the 2D analog of a WSM -- but the leading log approximation fails here and more analysis is needed \cite{FritzGrapheneConductivity, KashubaGraphene}.

\noindent {\it The Model.-}
In a WSM, the electronic dispersion about a Weyl node is generically of the form $
H_\text{Weyl}={\boldsymbol u}\cdot{\boldsymbol k}\,{ 1}+\sum_{a=1}^3{\boldsymbol v^a}\cdot{\boldsymbol k}\,\sigma^a
$, 
where $\sigma^a$ are the Pauli matrices. The velocities satisfy ${\boldsymbol v^1}\cdot ({\boldsymbol v^2}\times{\boldsymbol v^3})\neq 0$, and the Chern number $\pm 1$ (`chirality') associated with the Weyl node is ${\rm Sign}\left ( {\boldsymbol v^1}\cdot ({\boldsymbol v^2}\times{\boldsymbol v^3})\right )$. For simplicity, we shall drop the term proportional to identity and assume isotropic dispersion; relaxing this assumption should only produce small corrections. The Hamiltonian for a system of $N$ identically dispersing Weyl nodes (`flavors') with Coulomb interactions
and disorder may then be written as  $H  =  H_{0}+H_{I}+H_{D}$, with (repeated indices summed)
\begin{eqnarray}
H_{0} & = & \sum_a H_a = \sum_a \int_{\boldsymbol{k}}\psi_{\boldsymbol{k},a}^{\dagger}\left( \chi_av_{F}\boldsymbol{k\cdot\sigma}\right)\psi_{\boldsymbol{k},a}\nonumber\\
H_{I} & = & \frac{1}{2}\int_{\boldsymbol{k}_{1}\boldsymbol{k}_{2}\boldsymbol{q}}V(\boldsymbol{q})\psi_{\boldsymbol{k}_{2}-\boldsymbol{q},a\sigma}^{\dagger}\psi_{\boldsymbol{k}_{2},a\sigma}\psi_{\boldsymbol{k}_{1}+\boldsymbol{q},b\sigma^{\prime}}^{\dagger}\psi_{\boldsymbol{k}_{1}b\sigma\prime}\nonumber\\
H_{D} & = & \int_{\boldsymbol{r}} \sum_{a,b}\psi_{a}^{\dagger}(\boldsymbol{r})U(\boldsymbol{r})\psi_b(\boldsymbol{r}) \label{eq:disorderterm}
\end{eqnarray}
where $\psi_{\bk,a}$ is a two-component spinor in the (pseudo)spin indices $\sigma, \sigma'$, $a,b=1\dots N$ index the flavors, $v_F$ is the Fermi velocity, which we set to unity, 
$\chi_a = \pm1$ is the chirality of the $a^{th}$ Weyl node,
$V(\boldsymbol{q})=\frac{4\pi e^{2}}{\varepsilon q^{2}}$
describes the Coulomb interaction in a material with dielectric constant
$\varepsilon$, $U(\boldsymbol{r})$  is a random potential with white-noise correlations $\left\langle \left\langle U(\boldsymbol{r})U(\boldsymbol{r}^{\prime})\right\rangle \right\rangle = n_ \text{i} v_0^2 \delta(\boldsymbol{r} -\boldsymbol{r}')$
where $v_0$ characterizes the strength of the individual impurities and $n_ \text{i}$ their concentration, $\int_{\boldsymbol{k}}\equiv\int\frac{d^{3}k}{(2\pi)^{3}}$, and we have written $H_0$ assuming that the Fermi level is at the Weyl nodes, which is the only case studied in this Letter. Here and below we set $\hbar=k_B = |e| =1$, and define $\beta=1/T$.

\noindent {\it Conductivity with Interactions.-}\label{sec:interactions}
Critical systems -- such as graphene and the WSM  at neutrality-- are  exceptions to the  rule that disorder is essential for a finite conductivity, since they support current-carrying states in which particles and holes transport charge with no net momentum by moving exactly opposite to each other. In contrast to conventional finite-momentum charge transport, such deviations from equilibrium  {can} relax in the presence of interactions alone, leading to a finite conductivity.

We study transport in an interacting WSM by solving a QBE for the thermal distribution function of quasiparticle states. In doing so, it is convenient to first calculate the current from a single node (but interacting with all the nodes), before making the leap to the current carried by all $N$ nodes. We focus on a node with flavor $a$, which we take to have $\chi_a =1$. The single-quasiparticle states  are obtained by diagonalizing $H_a$: $\psi_{\bk,a} \rightarrow  U^\dagger\psi_{\bk,a} \equiv \gamma_{\bk,a} $,  $H_a\rightarrow U H_a U^{-1} = \int_{\bk}  \lambda v_F k \gamma^\dagger_{\bk\lambda a} \gamma_{\bk\lambda a}$, and are labeled by their helicity  $\lambda$ (the eigenvalue of $\sigma\cdot \hat{\boldsymbol{p}}$.) From now on we will suppress the index $a$. In general, operators corresponding to various transport properties are {\it not} diagonal in the helicity;  diagonal contributions correspond to motion of particles and holes in the applied field and may be characterized by appropriate distribution functions $f_\lambda(\bk, t) = \langle \gamma^\dagger_{ \bk \lambda} \gamma_{ \bk  \lambda}\rangle$, 
while the off-diagonal terms ($\gamma^\dagger_{\lambda\bk} \gamma_{-\lambda\bk}$) describe the motion of particle-hole pairs, which are finite-energy states. At particle-hole symmetry and for $\omega\ll T$,  contributions  of the latter to  transport are expected to be small, and we drop them forthwith. In this approximation, it is therefore sufficient to solve the QBE  for  quasiparticle distribution functions $f_\lambda(\bk,t)$, subject to an external force $\boldsymbol{F}$,
\be\label{eq:Boltzmann}
\left( \frac{\partial}{\partial t}  + \boldsymbol{F}\cdot\nabla_{\bk} \right) f_\lambda(\bk,t) = -w[f_\lambda(\bk,t)]
\ee
where $w$ is the rate at which quasiparticles scatter out of the state $(\lambda, \bk)$  at time $t$, and captures the effect of interactions.
Our goal will be to determine the steady-state form of the non-equilibrium quasiparticle distribution function. We will restrict ourselves to  linear response in $\boldsymbol{F}$, i.e. we assume that the deviation of $f_\lambda$ from equilibrium is small. The result is a  linear functional equation which may be recast as a variational problem. We   solve the latter approximately by identifying  `leading log' contributions,
 which dominate the relaxation of the observable under consideration. As mentioned, we assume that the constants that enter the solution of (\ref{eq:Boltzmann}) are renormalized to the energy scale of interest, namely $T$. 
 
 Neglecting particle-hole pair contributions,
the  current is $\boldsymbol{J}(t)=-\int_{\boldsymbol{k}}\langle\psi_{\boldsymbol{k}}^{\dagger}\boldsymbol{\sigma}\psi_{\boldsymbol{k}}\rangle_t=- \sum_{\lambda=\pm}\int_{\boldsymbol{k}}\lambda\hat{\boldsymbol{k}}f_{\lambda}(\boldsymbol{k},t)$.
For a weak applied electric field
$\boldsymbol{E}(t)$, the deviation of $f_{\lambda}(\boldsymbol{k},\omega)=\int dtf(\boldsymbol{k},t)e^{i\omega t}$ from the equilibrium distribution function $f_{\lambda}^{0}(k)=\left(1+e^{\lambda\beta k}\right)^{-1}$, and hence the conductivity $\sigma(\omega, T)$,
can be parametrized \footnote{We have used the fact that at particle-hole symmetry, the deviation from equilibrium is proportional to the sign of $\lambda$, and  $f^0_\lambda(k) = 1- f^0_{-\lambda}(k)$.} in terms of a dimensionless, isotropic function $g(k,\omega)$:  
  \begin{eqnarray}
f_{\lambda}(\boldsymbol{k},\omega)&=&2\pi\delta(\omega)f_{\lambda}^{0}(k) \nonumber\\ & &+\lambda\beta^{2}\boldsymbol{\hat{k}\cdot E}(\omega)[f_{\lambda}^{0}(k)f_{-\lambda}^{0}(k)]g(k,\omega)\label{eq:parametrization}\\
\sigma(\omega,T)&=&2\beta^{2}\int_{\boldsymbol{k}}\left[\frac{k_{x}^{2}}{k^{2}}[f_{+}^{0}(k)f_{-}^{0}(k)]g(k,\omega)\right]\label{eq:conductivity general}
\end{eqnarray}
It therefore remains only to determine the function $g(k,\omega)$, to which we now turn. Inserting (\ref{eq:parametrization}) into (\ref{eq:Boltzmann}), and working to linear order in $\boldsymbol E$, we find
$-\left(i\beta\omega g(k,\omega)+1\right)f_{+}^{0}(k)f_{-}^{0}(k)\hat{\boldsymbol{k}}={\hat{\mathcal{C}}}\left[g(k,\omega)\hat{\boldsymbol{k}}\right]\label{eq:QBE for g}$
 where ${\hat{\mathcal{C}}}$ is  the collision operator,
a linear functional of $g(k,\omega)\hat{\boldsymbol{k}}$ given in \ref{app:collision operator}. 
This is equivalent to the variational problem of extremizing the quadratic functional  \cite{GoswamiDiracTransport, KashubaGraphene, FritzGrapheneConductivity, LeadingLogTransport}
\be\label{eq:functional}\mathcal{Q}[g] &\equiv&  \int_{\bk}\left[ \frac12 g(k,\omega){\hat{\bk}}\cdot({\hat{\mathcal{C}}}[g(k,\omega)\hat{\bk}])\right.   \\& &\left.+f^0_+(k) f^0_-(k)\left(i\omega \frac{g^2(\omega,k)}{2} + g(\omega,k)\right)\right],  \nonumber \ee
in which we have rescaled all momenta and frequencies by $T$. A key simplification, known as the `leading log' approximation (LLA) stems from the power-law nature of the Coulomb interaction: as a result of this, logarithmically divergent  small-momentum scattering dominates  $\hat{\mathcal{C}}$. We may write  $\hat{\mathcal{C}} =\hat{\mathcal{C}}_0 +\hat{\mathcal{C}}_1$, which when thought of as linear functionals of $g \hat{\bk}$  have eigenvalues of $\mathcal{O}(\alpha^2\log\alpha)$ and $\mathcal{O}(\alpha^2)$, respectively.  
In the LLA we approximately optimize $\mathcal{Q}$ by choosing $g\hat{\boldsymbol{k}}$ in the space spanned by eigenstates of $\hat{\mathcal{C}}_0$; as shown in App. \ref{app:collision operator} the choice $g =  k\xi(\omega)$ yields
\be\label{eq:leadinglogQ}
\mathcal{Q}[k \xi(\omega)] &\approx& \frac{4}{\varepsilon^2}\left[i\omega [\xi(\omega)]^2\frac{7\pi^{4}}{30}+9 \xi(\omega)\zeta(3)\right]\nonumber\\ & &-\frac{4\pi^3}{9\varepsilon^2}[\xi(\omega)]^{2}N\alpha^{2}\ln\alpha^{-1},
\ee
optimized by $\xi(\omega)  = 
\frac{81\zeta(3)}{2\pi^{3}}\left({-i\omega\frac{21\pi}{10}+N\alpha^{2}\ln\alpha^{-1}}\right)^{-1}.$
Finally, we observe that the flipped chirality of half the nodes is unimportant as they all give the same contribution to $\sigma$; thus using the result for $\xi(\omega)$ in (\ref{eq:conductivity general}) and  multiplying by $N$ we find the result for $N$ nodes \footnote{As mentioned in the introduction, $\alpha$ and $v_F$ take on their renormalized values at temperature $T$} ,  
\be\label{eq:conductivityinteractions}
\sigma^{(N)}(\omega, T) = N\frac{e^{2}}{h}\frac{1.8}{-i\frac{\hbar\omega}{k_{B}T}6.6 + N\alpha^{2}\ln\alpha^{-1}}\left(\frac{k_{B}T}{\hbar v_{F}}\right)
\ee
Note that in the case of graphene, the LLA fails because the log divergence stems from a phase space effect due to enhanced scattering of collinear particles, which cannot relax a current. Thus, the eigenstates of $\hat{\mathcal{C}}_0$ do not contribute to the relaxation, which therefore occurs only via subleading, noncollinear scattering, i.e. $\hat{\mathcal{C}}_1$ \cite{FritzGrapheneConductivity, KashubaGraphene}. In 3D, $\hat{\mathcal{C}}_0$ includes noncollinear and thus  current-relaxing processes, so that the LLA analysis is sufficient \cite{LeadingLogTransport, GoswamiDiracTransport}.

In the dc limit, (\ref{eq:conductivityinteractions}) reduces to (\ref{eq:sigmadcint}), which we may rationalize in terms of the relaxation time approximation to the QBE. Here, we take $w[f_\lambda(\bk,t)]] = \frac{f_\lambda(\bk,t)}{\tau}$, and use a combination of dimensional analysis and a straightforward application of Fermi's golden rule to estimate the scattering rate as $\tau^{-1} \sim N\alpha^2 T$. From this, we find the conductivity to be similar to (\ref{eq:sigmadcint}), modulo logarithms. This  provides an estimate of the frequencies over which transport is collision-dominated and the preceding calculation is valid: in order for collisions to produce relaxation, we require $\omega\ll \tau^{-1}$, which occurs for $\hbar\omega/k_BT \ll N\alpha^2$.

%%%%%%%%%%%%%DISORDER%%%%%%%%%%%%%%%%
\noindent {\it Conductivity with impurities.-}\label{sec:disorder}
We turn now to the conductivity of the noninteracting, disordered system. We restrict  to the case of  scattering off random point impurities, characterized by $v_ \text{i}(\boldsymbol{r}) \sim v_0^2 \delta(\boldsymbol{r})$ and the locations of which we shall assume are uncorrelated, $\langle\langle \rho_ \text{i}(\boldsymbol{r}) \rho_ \text{i}(\boldsymbol{r}') \rangle\rangle \propto \delta(\boldsymbol{r} -\boldsymbol{r'})$. With these assumptions, we are led to  $H_D$ in (\ref{eq:disorderterm}) with $U(\br) \equiv \int d\br' v_ \text{i}(\br-\br')\rho_ \text{i}(\br')$.
As before, we first compute the conductivity for a single node. Assuming that the impurities are  sufficiently dilute that the Born approximation is valid, the quasiparticle lifetime due to impurity scattering from a single node is given by $\frac{1}{\tau(\omega)} = -2\text{Im}\Sigma^{{\text{ret}}}(\omega,\bk)$ where
$\Sigma_{\lambda}^{{\text{ret}}}(\omega,\bk) =n_ \text{i} v_0^2\int\frac{d^{3}k'}{(2\pi)^{3}}\mathcal{F}_{{\lambda\lambda'}}(\bk,\bk')G^{(0)}_{\lambda'}(\omega,\bk')
$ 
is the retarded self-energy,  $G_{\lambda}^{(0)}(\omega,\bk)=(\omega + i\delta-\lambda v_{F}k)^{-1}$ is the
Green's function for a noninteracting Weyl fermion with helicity $\lambda$, and the  form factor  from
the overlap of helicity eigenspinors, $\mathcal{F}_{\lambda\lambda'}(\bk,\bk')=\frac{1}{2}(1+\lambda\lambda'\cos\theta_{\bk\bk'})$ to leading order. 
 We find (see App. \ref{sec:scattertime}) $\frac{1}{\tau(\omega)}\equiv2\pi\gamma g(\omega)$, where
$g(\omega)=\frac{\omega^{2}}{2\pi^{2}v_{F}^{3}}$
is the density of states and  $\gamma = \frac{1}{2}n_ \text{i} v_0^2$ characterizes the strength of the impurity potential.

To evaluate the conductivity we use the Kubo formula, 
\be\label{eq:Kubodef}
\sigma(\omega, T) &=&
  -\frac{1}{\omega} \lim_{q\rightarrow0}   \text{Im} \,\Pi_{xx}^\text{ret}(\omega, |\bq|)
\ee
where  $\Pi_{\mu\nu}^\text{ret}(\omega, \bq)$ is the retarded response function which  for a system of linear dimension $L$ is defined to be
\be
\Pi^\text{ret}_{\mu\nu}(\omega, \bq) = -\frac{i}{L^3}\int_0^\infty dt e^{i\omega t}\left\langle \left[{J}_\mu(-\bq,t), {J}_\nu (\bq,0) \right]\right\rangle, 
\ee
 with $x_\mu = (t, \boldsymbol{r})$, $p_\mu = (\omega, \boldsymbol{p})$ and  $J_\mu = (-\psi^\dagger\psi, \boldsymbol{J})$.
From gauge invariance 
 $\Pi^\text{ret}_{\mu\nu}(\omega, \boldsymbol{q}) = \Pi^\text{ret}(\omega, |\boldsymbol{q}|) \left(\delta_{\mu\nu} - \frac{q_\mu q_\nu}{q^2} \right)$, so that (suppressing $\bq=0$),  
$\sigma(\omega, T) =
   -\frac{1}{\omega} \text{Im} \,\Pi^\text{ret}(\omega)
   = -\frac{1}{3\omega}
   \text{Im}\Pi^\text{ret}_{\mu\mu}(\omega)$.  
   Some algebra yields (see App. \ref{app:Kubo})
 \begin{eqnarray}\label{eq:Kubo}
\sigma(\omega,T)&=&\frac{4}{3}e^{2}v_{F}^{2}\int\frac{d\epsilon}{2\pi}\frac{\left[f_{T}(\epsilon)-f_{T}(\epsilon+\omega)\right]}{\omega} \\& & \times \sum_{{\lambda,\lambda'}}\int\frac{d^{3}k}{(2\pi)^{3}}\text{Im}G^{\text{ret}}_\lambda(\epsilon+\omega,k)\text{Im}G^{\text{ret}}_{\lambda'}(\epsilon,k).\nonumber%\\
\end{eqnarray}
where $f_{T}(\omega)=[e^{\omega/T}+1]^{-1}$ is
the Fermi-Dirac function and  we have used the retarded helicity-basis Green's function dressed with disorder
lines, $G_{\lambda}^{\text{ret}}(\omega,\bk)=[\omega-\lambda v_{F}k+i/2\tau(\omega)]^{-1}$.
After a tedious calculation, we may write
$\sigma(\omega,{T})=\frac{e^{2}v_{F}^{2}}{h\gamma}\mathcal{J}(\homega,\hat{T})\label{eq:conduniversal}
$
 where $\hat{T}=T/\omega_{0}$, $\homega=\omega/\omega_{0}$, so that $f_{T}(\omega)=f_{\hat{T}}(\homega)$,
$\omega_{0}=2\pi v_{F}^{3}/\gamma$ is a characteristic scale set by the disorder strength, and
$
\mathcal{J}(\homega,\hat{T})=\frac{4}{3}\int\frac{d\hepsilon}{2\pi}\frac{\left[f_{{\hat{T}}}(\hepsilon)-f_{{\hat{T}}}(\hepsilon+\homega)\right]}{\homega}\mathcal{I}(\hepsilon+\homega,\hepsilon)\label{eq:dimlessfunc}
$ 
with $\mathcal{I}$ a complicated rational function shown in App. \ref{app:Kubo}. 

\begin{figure}
\includegraphics[width=\columnwidth]{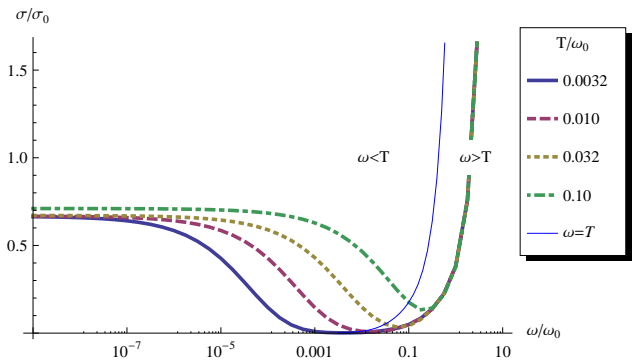}
\caption{\label{fig:condfreq}Frequency-dependent conductivity of a single Weyl node with disorder (constants defined in the text.)} \end{figure}

In our model  disorder can scatter between nodes, so $1/\tau(\omega)$ acquires a factor of $N$ when $N> 1$; in common with the interacting case, $\sigma$ also has an overall prefactor of $N$. From
these it is easy to show that for $N$ nodes,
\be\label{eq:Nnodecond}
\sigma^{(N)}(\omega, T) &=& \frac{e^2 v_F^2}{h\gamma} \mathcal{J}\left(N\frac{\omega}{\omega_0}, N\frac{T}{\omega_0}\right),
\ee
which is identical to the $N=1$ result (Fig. \ref{fig:condfreq}) upon rescaling $\omega_0 \rightarrow \omega_0/N$.

%%%%%%start figure
%%%%%%%%end  figure

While in general we integrate (\ref{eq:Nnodecond}) numerically, in certain  limits an  analytic treatment is feasible.
For $\omega\ll T$,  $f_{\hat{T}}(\hepsilon) - f_{\hat{T}}(\hepsilon+\homega) \approx -\homega f'(\hepsilon)$. Expanding $\mathcal{I}$ in powers of  $\homega$ and resumming only terms dominant as $\hepsilon\rightarrow 0$, we recover the result of  Burkov and Balents (Eq. (15) of \cite{WeylMultiLayer}): namely, a Drude-like response with a width vanishing as $NT^2$, and a finite dc limit of $\frac{2e^2 v_F^2}{3h\gamma}$. 

In the opposite limit, $T\rightarrow 0$ at finite $\omega$ we may replace the Fermi functions by step functions, which yields
 \be
\sigma^{(N)}(\omega) \approx
N\frac{e^2}{12 h} \frac{\omega}{v_F}  \left[1 - \frac{16N \gamma \omega}{15\pi^2 v_F^3}  +\mathcal{O}\left( \frac{N^2\omega^2}{\omega_0^2} \right)\right]\,\,\ee
The leading term is universal and independent of disorder, and is simply  $\sigma_0^{(N)}$. Both regimes are captured in Fig. \ref{fig:condfreq}, 
which shows $\sigma(\omega,T)$ for $\omega\lesssim\omega_0$, beyond which the Born approximation is insufficient.

\begin{figure}
\includegraphics[width=\columnwidth]{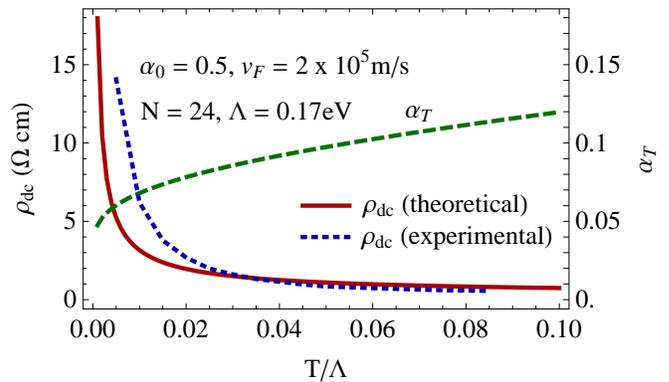}
\caption{\label{fig:combplots} $\rho_\text{dc} = \sigma_\text{dc}^{-1}$ and $\alpha_T$ (defined in the text) for the inset parameter values 
compared to experimental data from \cite{WeylResistivityMaeno}.} \end{figure}

\noindent{\it Experiments.-} 
In \cite{WeylResistivityMaeno} the dc resistivity of polycrystalline Y$_{2}$Ir$_{2}$O$_{7}$ was found to vary with temperature as
$\rho_{dc}T\approx130\Omega\cdot\mbox{cm}\cdot\mbox{K}$ over $10\text{K}\lesssim T\lesssim170\text{K}$, which  is reminiscent of our result with interactions (\ref{eq:sigmadcint}). Accordingly, we compare this data with a model of a clean WSM with $N=24$ \cite{PyrochloreWeyl}, as shown in Fig. \ref{fig:combplots}. We find rather good agreement with experimental data  
for physically reasonable parameter choices, shown inset. Very recently, transport in single crystals of another pyrochlore iridate, Eu$_2$Ir$_2$O$_7$, has been studied \cite{EuIridateExperiments}  under pressure for $2\text{K}\lesssim T \lesssim 300\text{K}$; at low pressures$\sim2.06-6.06\,\text{GPa}$, $\rho_\text{dc}(T)$ resembles Fig.~\ref{fig:combplots}, consistent with WSM behavior.

\noindent {\it Conclusions.-} The conductivity of WSMs thus exhibits a rich variety of behavior on varying frequency and temperature, in both the interacting clean and noninteracting disordered limits, as shown in Figs. \ref{fig:condfreq} and \ref{fig:combplots}. In particular, its nontrivial dependence on $N$ is sensitive to the strength of the interactions; with just disorder, we find a striking difference between the $\omega\ll T$ and $\omega\gg T$ regimes, with the $T\rightarrow 0$ ac response dominated by a universal, disorder-independent contribution. While the limited existing dc conductivity data on the candidate iridates 
broadly agrees with our  theory in the clean limit, we caution that more dc and ac conductivity measurements on single crystals with controlled disorder are required to make a rigorous comparison. Theoretically, the interplay of disorder and interactions, and corrections to the isotropic node approximation still need to be considered. In particular, it would be striking if the distinct behavior of the disordered system across the different frequency regimes survives the inclusion of interactions; these questions we leave open for the future.

{We thank P. Goswami, L. Fritz, S.L. Sondhi, S. Ryu and  A. Turner for discussions, and acknowledge funding from the Simons Foundation (SAP) and the Laboratory Directed Research and Development Program of LBNL under US DOE Contract DE-AC02-05CH11231 (AV).}

%%%%%%%%%%%%%%APPENDIX STARTS HERE%%%%%%%%%%%%%%%%%

\appendix
\section{The Collision Operator and the `Leading Log' Approximation\label{app:collision operator}}

In order to define the collision operator ${\hat{\mathcal{C}}}$, introduced in the main text, we need to determine the tree-level scattering amplitudes due to the interaction and how they couple the quasiparticle distributions at different momenta.
An additional complication stems from the fact that of the $N$ Weyl nodes, $N/2$ have  chirality $-1$. It is easily verified that the single-particle dispersion of the $a^\text{th}$ Weyl node is given by $E_{\lambda,a}(\boldsymbol{k})= \chi_a\lambda v_F k$, so that for $\chi_a =-1$  the eigenspinors of $H_a$ with eigenvalues $\pm v_F k$ are interchanged relative to $\chi_a =+1$. For convenience, we define $\epsilon_a = \chi_a \lambda$, so that the energy is $E_{a}(\boldsymbol{k})= \epsilon_a v_F k$ with $\epsilon_a = \pm 1$. As the Coulomb interaction is instantaneous, the collision operator is diagonal in frequency.  $H_{I}$ is then given by
\begin{eqnarray}
H_{I} & = & \frac{1}{2}\int_{\boldsymbol{k}_{1}\boldsymbol{k}_{2}\boldsymbol{q}}T^{\chi_a\chi_b}_{\epsilon_{1}\epsilon_{2}\epsilon_{3}\epsilon_{4}}(\boldsymbol{k}_{1}\boldsymbol{k}_{2}\boldsymbol{q})\times\nonumber \\
 &  & \gamma_{\epsilon_{4}b}^{\dagger}(\boldsymbol{k}_{1}+\boldsymbol{q})\gamma_{\epsilon_{3}a}^{\dagger}(\boldsymbol{k}_{2}-\boldsymbol{q})\gamma_{\epsilon_{2}a}(\boldsymbol{k}_{2})\gamma_{\epsilon_{1}b}(\boldsymbol{k}_{1})
\end{eqnarray}
 where
\begin{eqnarray}
T^{\chi_a\chi_b}_{\epsilon_{1}\epsilon_{2}\epsilon_{3}\epsilon_{4}}(\boldsymbol{k}_{1}\boldsymbol{k}_{2}\boldsymbol{q})&=&\frac{V(\boldsymbol{q})}{2}\left[U_{\chi_b}^{\dagger}({\boldsymbol{k}_{1}+\boldsymbol{q}})U_{\chi_b}({\boldsymbol{k}_{1}})\right]_{\epsilon_{4}\epsilon_{1}}\times\nonumber\\& &\left[U_{\chi_a}^{\dagger}({\boldsymbol{k}_{2}-\boldsymbol{q}})U_{\chi_a}(\boldsymbol{k}_{2})\right]_{\epsilon_{3}\epsilon_{1}}
\end{eqnarray}
is the scattering amplitude for two-particle scattering from $\{(b,\epsilon_{1},\boldsymbol{k}_{1}),(a,\epsilon_{2},\boldsymbol{k}_{2})\}$
to $\{(a,\epsilon_{3},\boldsymbol{k}_{2}-\boldsymbol{q}),(b,\epsilon_{4},\boldsymbol{k}_{1}+\boldsymbol{q})\}$
and in addition depends on the chiralities of the initial and final nodes. This dependence comes from the fact that the scattering amplitude depends explicitly on the overlap of the initial and final spinors, which in the $\epsilon$-representation introduced above depends on their chiralities.  We observe  here that in the ${q} \rightarrow 0$ limit (which, as we will shortly see, dominates the current relaxation) $T^{++}_{\epsilon_1\epsilon_2\epsilon_3\epsilon_4} = T^{+-}_{\epsilon_1\epsilon_2\epsilon_3\epsilon_4}$, corresponding to the  fact that $q\rightarrow0$ scattering is indifferent to the chirality, so the fact that half the Weyl nodes have a chirality opposing the one under consideration is unimportant.

Using Fermi's golden rule,
it is now straightforward to show that the collision operator is \begin{widetext}
\begin{eqnarray}
  {\hat{\mathcal{C}}}\left[g(k,\omega)\hat{\boldsymbol{k}}\right]
= & -2\pi & \intop_{\boldsymbol{k}^{\prime},\boldsymbol{q}}\Bigg\{\delta\left(k-k^{\prime}-\left|\boldsymbol{k}+\boldsymbol{q}\right|+\left|\boldsymbol{k}^{\prime}-\boldsymbol{q}\right|\right)f_{-}^{0}(k)f_{+}^{0}(k^{\prime})f_{+}^{0}(\left|\boldsymbol{k}+\boldsymbol{q}\right|)f_{-}^{0}(\left|\boldsymbol{k}^{\prime}-\boldsymbol{q}\right|)R_{1}\left(\boldsymbol{k},\boldsymbol{k}^{\prime},\boldsymbol{q}\right)\times\nonumber \\
 &  & \left[\frac{\boldsymbol{k}}{k}g(k,\omega)-\frac{\boldsymbol{k}^{\prime}}{k^{\prime}}g(k^{\prime},\omega)-\frac{\boldsymbol{k}+\boldsymbol{q}}{\left|\boldsymbol{k}+\boldsymbol{q}\right|}g(\left|\boldsymbol{k}+\boldsymbol{q}\right|,\omega)+\frac{\boldsymbol{k}^{\prime}-\boldsymbol{q}}{\left|\boldsymbol{k}^{\prime}-\boldsymbol{q}\right|}g(\left|\boldsymbol{k}^{\prime}-\boldsymbol{q}\right|,\omega)\right]+\nonumber \\
 &  & \delta\left(k+k^{\prime}-\left|\boldsymbol{k}+\boldsymbol{q}\right|-\left|\boldsymbol{k}^{\prime}-\boldsymbol{q}\right|\right)f_{-}^{0}(k)f_{-}^{0}(k^{\prime})f_{+}^{0}(\left|\boldsymbol{k}+\boldsymbol{q}\right|)f_{+}^{0}(\left|\boldsymbol{k}^{\prime}-\boldsymbol{q}\right|)R_{2}\left(\boldsymbol{k},\boldsymbol{k}^{\prime},\boldsymbol{q}\right)\times\nonumber \\
 &  & \left[\frac{\boldsymbol{k}}{k}g(k,\omega)+\frac{\boldsymbol{k}^{\prime}}{k^{\prime}}g(k^{\prime},\omega)-\frac{\boldsymbol{k}+\boldsymbol{q}}{\left|\boldsymbol{k}+\boldsymbol{q}\right|}g(\left|\boldsymbol{k}+\boldsymbol{q}\right|,\omega)-\frac{\boldsymbol{k}^{\prime}-\boldsymbol{q}}{\left|\boldsymbol{k}^{\prime}-\boldsymbol{q}\right|}g(\left|\boldsymbol{k}^{\prime}-\boldsymbol{q}\right|,\omega)\right]\Bigg\}
\end{eqnarray}
 where
\begin{eqnarray}
 R_{1}\left(\boldsymbol{k},\boldsymbol{k}^{\prime},\boldsymbol{q}\right) &=& \left|T^{++}_{+--+}(\boldsymbol{k},\boldsymbol{k}^{\prime},\boldsymbol{q})-T^{++}_{+-+-}(\boldsymbol{k},\boldsymbol{k}^{\prime},\boldsymbol{k}^{\prime}-\boldsymbol{k}-\boldsymbol{q})\right|^{2}\nonumber\\ & &\,\,\,\,\,\,\,+\left(\frac{N}{2}-1\right)\left[\left|T^{++}_{+--+}(\boldsymbol{k},\boldsymbol{k}^{\prime},\boldsymbol{q})\right|^{2}+\left|T^{++}_{+-+-}(\boldsymbol{k},\boldsymbol{k}^{\prime},\boldsymbol{k}^{\prime}-\boldsymbol{k}-\boldsymbol{q})\right|^{2}\right]\nonumber\\& &\,\,\,\,\,\,\,+\left(\frac{N}{2}\right)\left[\left|T^{+-}_{+--+}(\boldsymbol{k},\boldsymbol{k}^{\prime},\boldsymbol{q})\right|^{2}+\left|T^{+-}_{+-+-}(\boldsymbol{k},\boldsymbol{k}^{\prime},\boldsymbol{k}^{\prime}-\boldsymbol{k}-\boldsymbol{q})\right|^{2}\right]
\end{eqnarray}
 and
\begin{equation}
R_{2}\left(\boldsymbol{k},\boldsymbol{k}^{\prime},\boldsymbol{q}\right)=\frac{1}{2}\left|T^{++}_{++++}(\boldsymbol{k},\boldsymbol{k}^{\prime},\boldsymbol{q})-T^{++}_{++++}(\boldsymbol{k},\boldsymbol{k}^{\prime},\boldsymbol{k}^{\prime}-\boldsymbol{k}-\boldsymbol{q})\right|^{2}+\left(\frac{N}{2}-1\right)\left|T^{++}_{++++}(\boldsymbol{k},\boldsymbol{k}^{\prime},\boldsymbol{q})\right|^{2}+\left(\frac{N}{2}\right)\left|T^{+-}_{++++}(\boldsymbol{k},\boldsymbol{k}^{\prime},\boldsymbol{q})\right|^{2}
\end{equation}
 \end{widetext}describe scattering between oppositely and identically
charged particles, respectively. The diagrams corresponding to scattering
processes are the same as in Ref.~\onlinecite{FritzGrapheneConductivity}.

Turning now to the $\hat{\mathcal{C}}$-dependent portion of the expression for the functional $\mathcal{Q}$, we may use the symmetries of ${\hat{\mathcal{C}}}$ under combinatorial exchanges of momenta to write (in rescaled variables $k/T, \omega/T$) the first term in $\mathcal{Q}$ as
\begin{widetext}\begin{eqnarray}\label{eq:fullformquadraticfunctional}
\int_{\bk} g {\hat{\mathcal{C}}} g = -\frac{2\pi}{8}\int_{\boldsymbol{k},\boldsymbol{k}^{\prime},\boldsymbol{q}}\Bigg\{\delta\left(k-k^{\prime}-\left|\boldsymbol{k}+\boldsymbol{q}\right|+\left|\boldsymbol{k}^{\prime}-\boldsymbol{q}\right|\right)f_{-}^{0}(k)f_{+}^{0}(k^{\prime})f_{+}^{0}(\left|\boldsymbol{k}+\boldsymbol{q}\right|)f_{-}^{0}(\left|\boldsymbol{k}^{\prime}-\boldsymbol{q}\right|)R_{1}\left(\boldsymbol{k},\boldsymbol{k}^{\prime},\boldsymbol{q}\right) & \times\nonumber \\
\left[\frac{\boldsymbol{k}}{k}g(k,\omega)-\frac{\boldsymbol{k}^{\prime}}{k^{\prime}}g(k^{\prime},\omega)-\frac{\boldsymbol{k}+\boldsymbol{q}}{\left|\boldsymbol{k}+\boldsymbol{q}\right|}g(\left|\boldsymbol{k}+\boldsymbol{q}\right|,\omega)+\frac{\boldsymbol{k}^{\prime}-\boldsymbol{q}}{\left|\boldsymbol{k}^{\prime}-\boldsymbol{q}\right|}g(\left|\boldsymbol{k}^{\prime}-\boldsymbol{q}\right|,\omega)\right]^{2} + \nonumber\\
\delta\left(k+k^{\prime}-\left|\boldsymbol{k}+\boldsymbol{q}\right|-\left|\boldsymbol{k}^{\prime}-\boldsymbol{q}\right|\right)f_{-}^{0}(k)f_{-}^{0}(k^{\prime})f_{+}^{0}(\left|\boldsymbol{k}+\boldsymbol{q}\right|)f_{+}^{0}(\left|\boldsymbol{k}^{\prime}-\boldsymbol{q}\right|)R_{2}\left(\boldsymbol{k},\boldsymbol{k}^{\prime},\boldsymbol{q}\right)\times\nonumber \\
  \left[\frac{\boldsymbol{k}}{k}g(k,\omega)+\frac{\boldsymbol{k}^{\prime}}{k^{\prime}}g(k^{\prime},\omega)-\frac{\boldsymbol{k}+\boldsymbol{q}}{\left|\boldsymbol{k}+\boldsymbol{q}\right|}g(\left|\boldsymbol{k}+\boldsymbol{q}\right|,\omega)-\frac{\boldsymbol{k}^{\prime}-\boldsymbol{q}}{\left|\boldsymbol{k}^{\prime}-\boldsymbol{q}\right|}g(\left|\boldsymbol{k}^{\prime}-\boldsymbol{q}\right|,\omega)\right]^2\Bigg\}
\end{eqnarray}\end{widetext}

We now specialize to a perturbation of the form $g(k,\omega) = k\xi(\omega)$, which is a zero mode of the $R_2$ processes.
 This leaves only the terms in (\ref{eq:fullformquadraticfunctional}) proportional to $R_1$.
We may represent the energy-conserving $\delta$-function as
\begin{eqnarray}
\int_{-\infty}^{\infty}\delta\left(k-\left|\boldsymbol{k}+\boldsymbol{q}\right|+\Omega\right)\delta\left(k^{\prime}-\left|\boldsymbol{k}^{\prime}-\boldsymbol{q}\right|+\Omega\right)d\Omega\nonumber
\end{eqnarray}
Next, we choose coordinates in which $\boldsymbol{q}$ is along $\zhat$ so that
\begin{eqnarray}
\delta\left(k-\left|\boldsymbol{k}+\boldsymbol{q}\right|+\Omega\right) &=\delta\left(k-\sqrt{k^{2}+q^{2}+2kq\cos\theta}+\Omega\right)\nonumber\\&=\frac{k+\Omega}{kq}\delta\left(\cos\theta-\frac{\Omega^{2}+2k\Omega-q^{2}}{2kq}\right)\Theta(k+\Omega)\nonumber
\end{eqnarray}
and similarly for $\boldsymbol{k'}$.
The angular  integrals over $\theta$ and $\theta'$ can now be done trivially, at the cost of leaving a complicated dependence on $\Omega$ in $k$, $k'$.
For $q<k,k^{\prime}$, the $\theta$ and $\theta^{\prime}$ integrals
give one if $-q<\Omega<q$ and 0 otherwise; we will show that it suffices to consider small $q$, so the knowledge of this regime is sufficient. With these observations, we have
\begin{widetext}
\begin{eqnarray}
\int_{\bk} g {\hat{\mathcal{C}}} g = -\frac{2\pi}{8}\int_{\boldsymbol{k},\boldsymbol{k}^{\prime},\boldsymbol{q}}\int_{-q}^{q}d\Omega\frac{k^{\prime}+\Omega}{k^{\prime}q}\frac{k+\Omega}{kq}f_{-}^{0}(k)f_{+}^{0}(k^{\prime})f_{+}^{0}(\left|\boldsymbol{k}+\boldsymbol{q}\right|)f_{-}^{0}(\left|\boldsymbol{k}^{\prime}-\boldsymbol{q}\right|)R_{1}\left(\boldsymbol{k},\boldsymbol{k}^{\prime},\boldsymbol{q}\right) & \times\nonumber \\
\left[\frac{\boldsymbol{k}}{k}g(k,\omega)-\frac{\boldsymbol{k}^{\prime}}{k^{\prime}}g(k^{\prime},\omega)-\frac{\boldsymbol{k}+\boldsymbol{q}}{\left|\boldsymbol{k}+\boldsymbol{q}\right|}g(\left|\boldsymbol{k}+\boldsymbol{q}\right|,\omega)+\frac{\boldsymbol{k}^{\prime}-\boldsymbol{q}}{\left|\boldsymbol{k}^{\prime}-\boldsymbol{q}\right|}g(\left|\boldsymbol{k}^{\prime}-\boldsymbol{q}\right|,\omega)\right]^{2}
\end{eqnarray}
\end{widetext}
The cross-helicity scattering rate $R_{1}$ contains terms of the form $V(q)^{2}$, which diverge as $1/q^{4}$ at small
 $q$ for direct scattering; the exchange contributions remain finite and do not contribute at leading order. The integral over the dummy $\Omega$ variable gives one power of $q$ in thenumerator and  substitution of our ansatz shows that the term in braces gives two more powers of $q$, so that at small
$q$, the integrand is proportional to $1/q$, which gives a logarithmic divergence. It is this logarithmically dominant contribution that is referred to in the text as $\hat{\mathcal{C}}_0$, whence we implicitly define $\hat{\mathcal{C}}_1 = \hat{\mathcal{C}} - \hat{\mathcal{C}}_0$.

Having identified the dominant contribution to the collision integral, we proceed as follows: first, we substitute  our ansatz into the expression above, and perform the integral over ${q}$; as this integral is logarithmically divergent in both the UV and the IR, we will need to regulate it appropriately.  The UV cutoff is the temperature $T$, as this is the largest  physical energy scale in the problem; in the IR, the divergence is avoided due to thermal screening of the Coulomb interaction which occurs at an energy scale set by both the coupling constant and the temperature, which we take to be $\alpha T$. With these choices, we may perform the integral over $q$, and then set $q=0$ in the remainder, leaving  integrals over $\Omega$, $k$ and $k'$ which may then be performed straightforwardly; upon including the parts of the functional $\mathcal{Q}$ that are independent of the collision operator, we arrive at the results quoted in the text for the LLA.

\section{Renormalization of couplings}\label{app:RG}
\begin{figure}
\includegraphics[width=0.7\columnwidth]{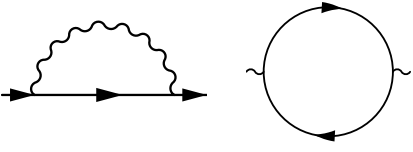}
\caption{\label{fig:RGdiagrams} One-loop diagrams that govern the flow of the marginal couplings $v_F$ and $\epsilon$.}
\end{figure}
As mentioned in the text, in order to make a meaningful comparision between the calculated conductivity and  experimental data at temperature $T$, we need to use values of the various parameters renormalized to the characteristic energy scale $k_BT$. In 
order to determine these,  we work with the Euclidean path integral
  $\mathcal{Z}[\boldsymbol{A}]=\int [DA_0 D\psi^\dagger D\psi] e^{-\mathcal{S}[\psi, A_0, \boldsymbol{A}]}$, written in terms of the gauge-fixed action
\begin{eqnarray}
\mathcal{S}[\psi, A_0,\boldsymbol{A}] &=& \int d\boldsymbol r d\tau\bigg[  \psi^\dagger_a(\boldsymbol r,\tau)[\partial_\tau-i|e|A_\tau(\boldsymbol r,\tau)\nonumber\\
& &+v_F\boldsymbol\sigma\cdot(i\boldsymbol\nabla+|e|\boldsymbol A/c)] \psi_a(\boldsymbol r,\tau)\nonumber\\
 & & + \frac{1}{2}\int_{\boldsymbol q}\int d\tau \frac{\varepsilon q^2}{4\pi}|A_\tau(\boldsymbol q,\tau)|^2\bigg]
\end{eqnarray}
which (as is easily verified) is equivalent to the Hamiltonian $H$ described earlier. Here,  $\boldsymbol A$ is a static external vector potential, 
while $A_\tau$ is a dynamical field that mediates the $e^2/(\varepsilon r)$ instantaneous Coulomb interaction between the electrons. We implement a momentum-space cutoff $k<\Lambda$, but leave the frequencies unrestricted so that the interaction always remains instantaneous; we then study the flow of $\mathcal{S}$  as we lower $\Lambda$. The action has three dimensionless parameters -- the Fermi velocity $v_F$, the electric charge $e$ and the dielectric constant $\varepsilon$ -- all of which are unrenormalized, and hence marginal, at  tree level. Of these, $e$ is further protected against renormalization to \emph{all orders} in perturbation theory by gauge invariance. However, $v_F$ and $\varepsilon$ have no such protection and at one loop they acquire corrections of $\mathcal O(\alpha\frac{\delta\Lambda}{\Lambda})$ when  $\Lambda$ is lowered to $\Lambda-\delta\Lambda$. The diagrams governing the flow of $v_F$ and $\varepsilon$ at this order in $\alpha$ are the one-loop self-energy  and the fermion (RPA) bubble  (see Fig.~\ref{fig:RGdiagrams}) diagrams respectively; performing the sum over flavors where appropriate, we find that the corrections are, respectively, independent of and proportional to $N$. Evaluation of the corresponding integrals leads to the flow equations

\be\label{eq:RGequations}
\frac{d v_F}{d\ell} = \frac{2}{3\pi} \alpha v_F,\,\,\,\frac{d\varepsilon}{d\ell}=\frac{N}{3\pi}\alpha\varepsilon\implies\frac{d \alpha}{d\ell} = -\frac{N+2}{3\pi} \alpha^2,
\ee
where $\ell = \log \Lambda$ is the usual RG scale parameter.
Integrating (\ref{eq:RGequations}) between an upper cutoff (which we leave unspecified for the time being, but will take to be much higher than any scale relevant to transport measurements, but lower than the energy at which band curvature corrections become significant) and the  temperature scale, $T$ gives the renormalized parameters used in the text.  These renormalized values of $v_F$ and $\alpha$ are to be used in the formula for $\sigma_\text{dc}$ [Eq.(1) of the main text] to compute the conductivity at temperature $T$, while $e$ retains its bare value due to gauge invariance. Representative results  are shown in Fig. \ref{fig:resistivityplot}.

We remark on an important distinction between the case of the Weyl semimetal in $d=3$ and its two-dimensional analog, graphene. In the case of graphene, while the electrons are restricted to two dimensions, their interaction continues to be of the $1/r$ form since the electromagnetic field lines are not thus constrained; in momentum space, this leads to a non-analytic form for the instantaneous interaction, $V(\boldsymbol{q})\sim 1/|\boldsymbol{q}|$. Since the RG procedure outlined above can only produce analytic corrections at any order, $\varepsilon$ thus remains unrenormalized. In $d=3$, we have an analytic $V(\boldsymbol{q})\sim 1/q^2$, and so $\varepsilon$ can and does receive corrections at one loop and beyond.

%%%%%%start figure
\begin{figure}
\includegraphics[width=\columnwidth]{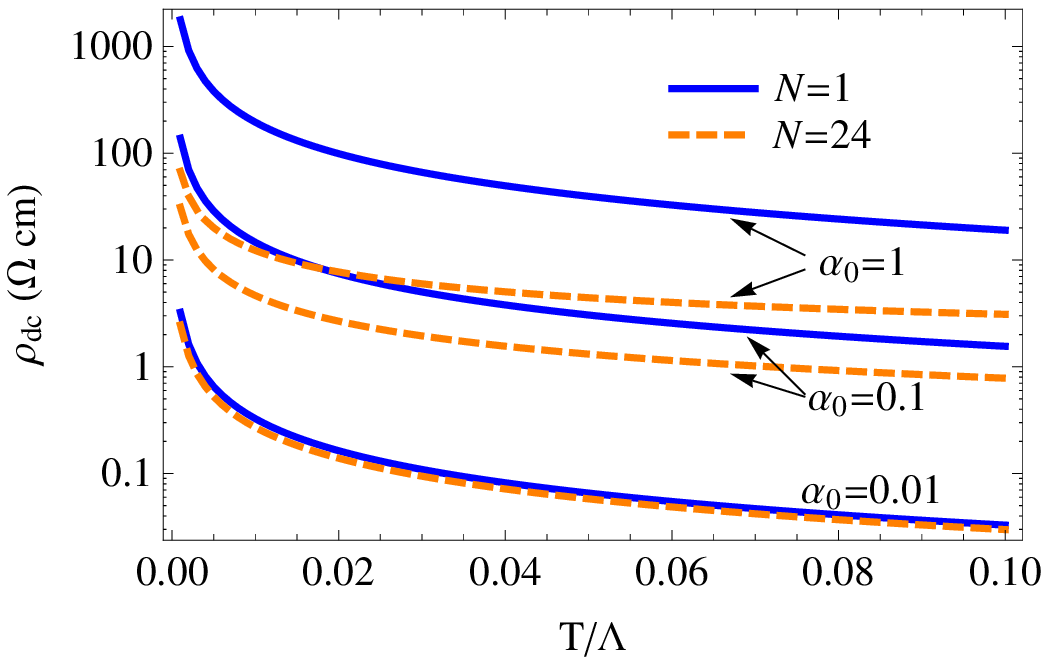}
\caption{\label{fig:resistivityplot}Resistivity as a function of temperature for $N$-node Weyl semimetals  with $N=1$ and $N=24$.}
\includegraphics[width=\columnwidth]{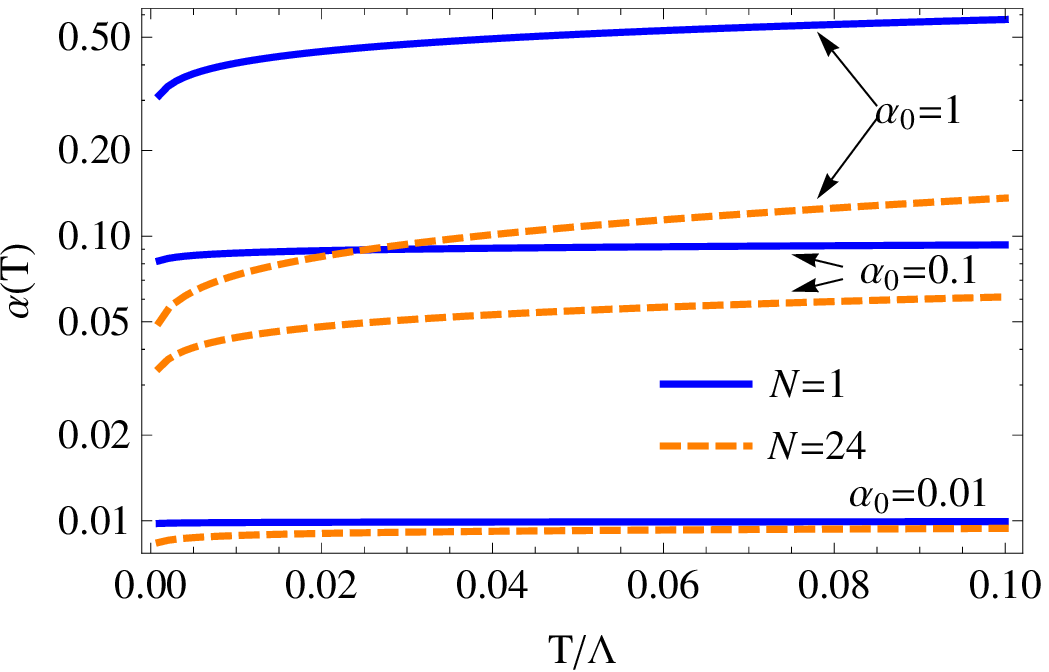}
\caption{\label{fig:resistivityplot}Renormalized fine-structure constant as a function of temperature for $N$-node Weyl semimetals  with $N=1$ and $N=24$.}
\end{figure}
%%%%%%%%end  figure

\section{Conductivity with Disorder}\label{sec:disorder}
\subsection{Scattering Time}\label{sec:scattertime}
The scattering time is given in terms of the imaginary part of the retarded self-energy by$
\frac{1}{\tau(\omega)} = -2\text{Im}\Sigma_{\lambda}^{{\text{ret}}}(\omega,\bk)$, where
\begin{align}
\text{Im}\Sigma_{\lambda}^{{\text{ret}}}(\omega,\bk)& =n_\text{i} v_0^2\text{Im}\int\frac{d^{3}k'}{(2\pi)^{3}}\mathcal{F}_{{\lambda\lambda'}}(\bk,\bk')G^{(0)}_{\lambda'}(\omega,\bk').
\end{align}
with all quantities defined as in the text. Imposing a  cutoff $|\bk| < \Omega/v_F$, performing the angular integration, and rewriting the negative-helicity Green's function by changing variables we have
\begin{align}
 \text{Im}\Sigma_{\lambda}^{{\text{ret}}}(\omega,\bk)
 & =\frac{n_\text{i} v_0^2}{4\pi^{2}v_{F}^{3}}\text{Im}\int_{-\Omega}^{\Omega}dx\,\frac{x^{2}}{\omega+i\delta-x}\nonumber\\
 & =-\frac{n_\text{i} v_0^2 \omega^2}{4\pi v_{F}^{3}},
\end{align} independent of $\lambda$,
whence $
\frac{1}{\tau(\omega)}\equiv2\pi\gamma g(\omega)$, where
$g(\omega)=\frac{\omega^{2}}{2\pi^{2}v_{F}^{3}}$
is the density of states and  $\gamma = \frac{1}{2}n_\text{i} v_0^2$. Note that  $\text{Re}\Sigma^{\text{ret}}(\omega,\bk)$ is formally divergent with the cutoff regularization used
here; however this can be absorbed into a renormalization of $v_F$ and we shall not discuss it further.
 
\subsection{Evaluation of the Kubo Response}\label{app:Kubo}

In order to evaluate the conductivity we require the retarded response function  $\Pi_{\mu\nu}^\text{ret}(\omega)$ which is obtained most simply by working in the Matsubara formalism  and analytically continuing the results to real frequency. The leading  contribution to  the Matsubara response  function $\Pi_{\mu\nu}$ corresponds to the particle-hole bubble with impurity-dressed propagators, and in terms of Matsubara Green's functions in the pseudospin basis takes the form
 \be
 \Pi_{\mu\nu} (i\omega_n) &=& \frac{e^2v_F^2}{\beta} \sum_m \int_{\bk} \text{Tr} \left[ \mathcal{G}(i(\omega_n+\epsilon_m), \bk) \sigma_\mu\right. \nonumber\\&& \,\,\,\,\,\,\,\,\,\,\,\, \times \left. \mathcal{G}(i\epsilon_m, \bk) \sigma_\nu\right]
 \ee
where $\mathcal{G}(i\omega_n, \bk) = (i\omega_n - \sigma\cdot \bk + i/2\tau(\omega_n))^{-1} $ and  the trace is over the pseudospin indices. Performing this trace  and summing over the spatial index $\mu$, we  obtain
\be
 \Pi_{\mu\mu} (i\omega_n) &=& \frac{2e^2v_F^2}{\beta} \sum_{{\substack{m\\ \lambda,\lambda'}}}\int_{\bk} \mathcal{G}_\lambda(i(\omega_n+\epsilon_m), \bk)  \mathcal{G}_{\lambda'}(i\epsilon_m, \bk)\nonumber\\\ee
where $\mathcal{G}_\lambda(i\omega_n, \bk) = (i\omega_n -\lambda v_F \bk + i/2\tau(\omega_n))^{-1}$ is the Matsubara Green's function, now  written in the helicity basis.
Performing the sum over Matsubara frequencies by standard methods, analytically continuing the result  to real frequencies via  $\omegan\rightarrow \omega+i\delta$ and using \be\label{eq:conductivitysum}
\sigma(\omega, T) &=&
   -\frac{1}{\omega} \text{Im} \,\Pi^\text{ret}(\omega)
   = -\frac{1}{3\omega}
   \text{Im}\Pi^\text{ret}_{\mu\mu}(\omega).
\ee
 we arrive at the result,
 \begin{eqnarray}\label{eq:Kubo}
\sigma(\omega,T)&=&\frac{4}{3}e^{2}v_{F}^{2}\int\frac{d\epsilon}{2\pi}\frac{\left[f_{T}(\epsilon)-f_{T}(\epsilon+\omega)\right]}{\omega} \nonumber\\& & \times \sum_{{\lambda,\lambda'}}\int\frac{d^{3}k}{(2\pi)^{3}}\text{Im}G^{\text{ret}}_\lambda(\epsilon+\omega,k)\text{Im}G^{\text{ret}}_{\lambda'}(\epsilon,k)\nonumber\\
\end{eqnarray}

In order to perform the integrations in (\ref{eq:Kubo}), it is convenient to write $\frac{1}{2\tau(\omega)}=\frac{\gamma}{2\pi v_{F}^{3}}\omega^{2}\equiv\frac{\omega^{2}}{\omega_{0}}$.
With this definition, we have
\begin{align}
\text{Im}G_{\lambda}^{\text{ret}}(\omega,\bk) & =\text{Im}\frac{1}{\omega-\lambda v_{F}k+i\frac{\omega^{2}}{\omega_{0}}} \nonumber\\ & =\frac{\omega^{2}}{\omega_{0}^{3}}\frac{1}{\left(\frac{v_{F}k}{\omega_{0}}-\lambda\frac{\omega}{\omega_{0}}\right)^{2}+\left(\frac{\omega}{\omega_{0}}\right)^{4}}
\end{align}
By a straightforward change of variable, followed by an angular integral  we can write the sum over helicities and the dimensionless frequency parameter $\homega_{i}=\omega_{i}/\omega_{0}$, for convenience
replacing $\epsilon+\omega$ and $\epsilon$ by $\omega_{i}$
\begin{equation}
\sum_{{\lambda_1,\lambda_2}}\int\frac{d^{3}k}{(2\pi)^{3}}\text{Im}G_{{\lambda_{1}}}^{\text{ret}}(\omega_{1},\bk)\text{Im}G_{{\lambda_{2}}}^{\text{ret}}(\omega_{2},\bk)  = \frac{1}{2\pi\gamma} \mathcal{I}(\homega_1,\homega_2)
\end{equation}
which serves to define the function $\mathcal{I}$, which contains a single integral over the dimensionless parameter $x = v_F |\bk|/\omega_0$.
Rewriting the Kubo formula and reinstating factors of $\hbar$ by dimensional
analysis we fin
we can express the conductivity in terms
of a single function  of rescaled temperature and frequency,
\begin{equation}
\sigma(\omega,{T})=\frac{e^{2}v_{F}^{2}}{h\gamma}\mathcal{J}(\homega,\hat{T})\label{eq:conduniversal}
\end{equation}
 where $\hat{T}=T/\omega_{0}$, $\homega=\omega/\omega_{0}$, so that $f_{T}(\omega)=f_{\hat{T}}(\homega)$,
$\omega_{0}=2\pi v_{F}^{3}/\gamma$ is a characteristic scale set by the strength of the
disorder, and
\begin{equation}
\mathcal{J}(\homega,\hat{T})=\frac{4}{3}\int\frac{d\hepsilon}{2\pi}\frac{\left[f_{{\hat{T}}}(\hepsilon)-f_{{\hat{T}}}(\hepsilon+\homega)\right]}{\homega}\mathcal{I}(\hepsilon+\homega,\hepsilon)\label{eq:dimlessfunc}.
\end{equation}
Numerical integration of this expression yields the conductivity
for all $T$, $\omega$, as shown in Figure 1 of the main text. and direct evaluation of the integrals yields an explicit form for $\mathcal{I}$,

\begin{widetext}
\begin{equation}
\mathcal{I}(\hepsilon+\homega,\hepsilon)=\frac{4\pi\hepsilon^{2}(\hepsilon+\homega)^{2}(2\hepsilon^{2}+2\hepsilon\homega+\homega^{2})(1+2\hepsilon^{2}+2\hepsilon\homega+\homega^{2})(2+2\hepsilon^{2}+2\hepsilon\homega+\homega^{2})}{(4\hepsilon^{4}+8\hepsilon^{3}\homega+\homega^{2}+8\hepsilon^{2}\homega^{2}+4\hepsilon\homega^{3}+\homega^{4})(4\hepsilon^{4}+8\hepsilon^{3}\homega+\homega^{2}+\homega^{4}+\hepsilon^{2}(4+8\homega^{2})+4\hepsilon(\homega+\homega^{3}))}.
\end{equation}
\end{widetext}

\bibliography{Weyl_References}

%merlin.mbs apsrev4-1.bst 2010-07-25 4.21a (PWD, AO, DPC) hacked
%Control: key (0)
%Control: author (8) initials jnrlst
%Control: editor formatted (1) identically to author
%Control: production of article title (-1) disabled
%Control: page (0) single
%Control: year (1) truncated
%Control: production of eprint (0) enabled
\begin{thebibliography}{20}%
\makeatletter
\providecommand \@ifxundefined [1]{%
 \@ifx{#1\undefined}
}%
\providecommand \@ifnum [1]{%
 \ifnum #1\expandafter \@firstoftwo
 \else \expandafter \@secondoftwo
 \fi
}%
\providecommand \@ifx [1]{%
 \ifx #1\expandafter \@firstoftwo
 \else \expandafter \@secondoftwo
 \fi
}%
\providecommand \natexlab [1]{#1}%
\providecommand \enquote  [1]{``#1''}%
\providecommand \bibnamefont  [1]{#1}%
\providecommand \bibfnamefont [1]{#1}%
\providecommand \citenamefont [1]{#1}%
\providecommand \href@noop [0]{\@secondoftwo}%
\providecommand \href [0]{\begingroup \@sanitize@url \@href}%
\providecommand \@href[1]{\@@startlink{#1}\@@href}%
\providecommand \@@href[1]{\endgroup#1\@@endlink}%
\providecommand \@sanitize@url [0]{\catcode `\\12\catcode `\$12\catcode
  `\&12\catcode `\#12\catcode `\^12\catcode `\_12\catcode `\%12\relax}%
\providecommand \@@startlink[1]{}%
\providecommand \@@endlink[0]{}%
\providecommand \url  [0]{\begingroup\@sanitize@url \@url }%
\providecommand \@url [1]{\endgroup\@href {#1}{\urlprefix }}%
\providecommand \urlprefix  [0]{URL }%
\providecommand \Eprint [0]{\href }%
\providecommand \doibase [0]{http://dx.doi.org/}%
\providecommand \selectlanguage [0]{\@gobble}%
\providecommand \bibinfo  [0]{\@secondoftwo}%
\providecommand \bibfield  [0]{\@secondoftwo}%
\providecommand \translation [1]{[#1]}%
\providecommand \BibitemOpen [0]{}%
\providecommand \bibitemStop [0]{}%
\providecommand \bibitemNoStop [0]{.\EOS\space}%
\providecommand \EOS [0]{\spacefactor3000\relax}%
\providecommand \BibitemShut  [1]{\csname bibitem#1\endcsname}%
\let\auto@bib@innerbib\@empty
%</preamble>
\bibitem [{\citenamefont {Geim}\ and\ \citenamefont
  {Novoselov}(2007)}]{GeimGraphene}%
  \BibitemOpen
  \bibfield  {author} {\bibinfo {author} {\bibfnamefont {A.~K.}\ \bibnamefont
  {Geim}}\ and\ \bibinfo {author} {\bibfnamefont {K.~S.}\ \bibnamefont
  {Novoselov}},\ }\href@noop {} {\bibfield  {journal} {\bibinfo  {journal} {Nat
  Mater}\ }\textbf {\bibinfo {volume} {6}},\ \bibinfo {pages} {183} (\bibinfo
  {year} {2007})}\BibitemShut {NoStop}%
\bibitem [{\citenamefont {Herring}(1937)}]{Herring}%
  \BibitemOpen
  \bibfield  {author} {\bibinfo {author} {\bibfnamefont {C.}~\bibnamefont
  {Herring}},\ }\href {\doibase 10.1103/PhysRev.52.361} {\bibfield  {journal}
  {\bibinfo  {journal} {Phys. Rev.}\ }\textbf {\bibinfo {volume} {52}},\
  \bibinfo {pages} {361} (\bibinfo {year} {1937})}\BibitemShut {NoStop}%
\bibitem [{\citenamefont {Abrikosov}\ and\ \citenamefont
  {Beneslavskii}(1971)}]{Abrikosov}%
  \BibitemOpen
  \bibfield  {author} {\bibinfo {author} {\bibfnamefont {A.~A.}\ \bibnamefont
  {Abrikosov}}\ and\ \bibinfo {author} {\bibfnamefont {S.~D.}\ \bibnamefont
  {Beneslavskii}},\ }\href@noop {} {\bibfield  {journal} {\bibinfo  {journal}
  {J. Low Temp. Phys.}\ }\textbf {\bibinfo {volume} {5}},\ \bibinfo {pages}
  {141} (\bibinfo {year} {1971})}\BibitemShut {NoStop}%
\bibitem [{\citenamefont {Nielsen}\ and\ \citenamefont
  {Ninomiya}(1983)}]{NielsenABJ}%
  \BibitemOpen
  \bibfield  {author} {\bibinfo {author} {\bibfnamefont {H.}~\bibnamefont
  {Nielsen}}\ and\ \bibinfo {author} {\bibfnamefont {M.}~\bibnamefont
  {Ninomiya}},\ }\href {\doibase 10.1016/0370-2693(83)91529-0} {\bibfield
  {journal} {\bibinfo  {journal} {Physics Letters B}\ }\textbf {\bibinfo
  {volume} {130}},\ \bibinfo {pages} {389 } (\bibinfo {year}
  {1983})}\BibitemShut {NoStop}%
\bibitem [{\citenamefont {Wan}\ \emph {et~al.}(2011)\citenamefont {Wan},
  \citenamefont {Turner}, \citenamefont {Vishwanath},\ and\ \citenamefont
  {Savrasov}}]{PyrochloreWeyl}%
  \BibitemOpen
  \bibfield  {author} {\bibinfo {author} {\bibfnamefont {X.}~\bibnamefont
  {Wan}}, \bibinfo {author} {\bibfnamefont {A.~M.}\ \bibnamefont {Turner}},
  \bibinfo {author} {\bibfnamefont {A.}~\bibnamefont {Vishwanath}}, \ and\
  \bibinfo {author} {\bibfnamefont {S.~Y.}\ \bibnamefont {Savrasov}},\ }\href
  {\doibase 10.1103/PhysRevB.83.205101} {\bibfield  {journal} {\bibinfo
  {journal} {Phys. Rev. B}\ }\textbf {\bibinfo {volume} {83}},\ \bibinfo
  {pages} {205101} (\bibinfo {year} {2011})}\BibitemShut {NoStop}%
\bibitem [{\citenamefont {Yang}\ \emph {et~al.}(2011)\citenamefont {Yang},
  \citenamefont {Lu},\ and\ \citenamefont {Ran}}]{RanQHWeyl}%
  \BibitemOpen
  \bibfield  {author} {\bibinfo {author} {\bibfnamefont {K.-Y.}\ \bibnamefont
  {Yang}}, \bibinfo {author} {\bibfnamefont {Y.-M.}\ \bibnamefont {Lu}}, \ and\
  \bibinfo {author} {\bibfnamefont {Y.}~\bibnamefont {Ran}},\ }\href {\doibase
  10.1103/PhysRevB.84.075129} {\bibfield  {journal} {\bibinfo  {journal} {Phys.
  Rev. B}\ }\textbf {\bibinfo {volume} {84}},\ \bibinfo {pages} {075129}
  (\bibinfo {year} {2011})}\BibitemShut {NoStop}%
\bibitem [{\citenamefont {Burkov}\ and\ \citenamefont
  {Balents}(2011)}]{WeylMultiLayer}%
  \BibitemOpen
  \bibfield  {author} {\bibinfo {author} {\bibfnamefont {A.~A.}\ \bibnamefont
  {Burkov}}\ and\ \bibinfo {author} {\bibfnamefont {L.}~\bibnamefont
  {Balents}},\ }\href {\doibase 10.1103/PhysRevLett.107.127205} {\bibfield
  {journal} {\bibinfo  {journal} {Phys. Rev. Lett.}\ }\textbf {\bibinfo
  {volume} {107}},\ \bibinfo {pages} {127205} (\bibinfo {year}
  {2011})}\BibitemShut {NoStop}%
\bibitem [{\citenamefont {{Aji}}(2011)}]{AjiABJAnomaly}%
  \BibitemOpen
  \bibfield  {author} {\bibinfo {author} {\bibfnamefont {V.}~\bibnamefont
  {{Aji}}},\ }\href@noop {} {\bibfield  {journal} {\bibinfo  {journal} {ArXiv
  e-prints}\ } (\bibinfo {year} {2011})},\ \Eprint
  {http://arxiv.org/abs/1108.4426} {arXiv:1108.4426 [cond-mat.str-el]}
  \BibitemShut {NoStop}%
\bibitem [{\citenamefont {Turner}\ \emph {et~al.}(2010)\citenamefont {Turner},
  \citenamefont {Zhang}, \citenamefont {Mong},\ and\ \citenamefont
  {Vishwanath}}]{TurnerAV}%
  \BibitemOpen
  \bibfield  {author} {\bibinfo {author} {\bibfnamefont {A.}~\bibnamefont
  {Turner}}, \bibinfo {author} {\bibfnamefont {Y.}~\bibnamefont {Zhang}},
  \bibinfo {author} {\bibfnamefont {R.}~\bibnamefont {Mong}}, \ and\ \bibinfo
  {author} {\bibfnamefont {A.}~\bibnamefont {Vishwanath}},\ }\href@noop {}
  {\bibfield  {journal} {\bibinfo  {journal} {ArXiv e-prints}\ } (\bibinfo
  {year} {2010})},\ \Eprint {http://arxiv.org/abs/1010.4335} {arXiv:1010.4335
  [cond-mat.str-el]} \BibitemShut {NoStop}%
\bibitem [{\citenamefont {Witczak-Krempa}\ and\ \citenamefont
  {Kim}(2011)}]{Kim}%
  \BibitemOpen
  \bibfield  {author} {\bibinfo {author} {\bibfnamefont {W.}~\bibnamefont
  {Witczak-Krempa}}\ and\ \bibinfo {author} {\bibfnamefont {Y.~B.}\
  \bibnamefont {Kim}},\ }\href@noop {} {\bibfield  {journal} {\bibinfo
  {journal} {ArXiv e-prints}\ } (\bibinfo {year} {2011})},\ \Eprint
  {http://arxiv.org/abs/1105.6108} {arXiv:1105.6108 [cond-mat.str-el]}
  \BibitemShut {NoStop}%
\bibitem [{\citenamefont {{Xu}}\ \emph {et~al.}(2011)\citenamefont {{Xu}},
  \citenamefont {{Weng}}, \citenamefont {{Wang}}, \citenamefont {{Dai}},\ and\
  \citenamefont {{Fang}}}]{FangChernSemiMetal}%
  \BibitemOpen
  \bibfield  {author} {\bibinfo {author} {\bibfnamefont {G.}~\bibnamefont
  {{Xu}}}, \bibinfo {author} {\bibfnamefont {H.}~\bibnamefont {{Weng}}},
  \bibinfo {author} {\bibfnamefont {Z.}~\bibnamefont {{Wang}}}, \bibinfo
  {author} {\bibfnamefont {X.}~\bibnamefont {{Dai}}}, \ and\ \bibinfo {author}
  {\bibfnamefont {Z.}~\bibnamefont {{Fang}}},\ }\href@noop {} {\bibfield
  {journal} {\bibinfo  {journal} {ArXiv e-prints}\ } (\bibinfo {year}
  {2011})},\ \Eprint {http://arxiv.org/abs/1106.3125} {arXiv:1106.3125
  [cond-mat.mes-hall]} \BibitemShut {NoStop}%
\bibitem [{\citenamefont {Yanagishima}\ and\ \citenamefont
  {Maeno}(2001)}]{WeylResistivityMaeno}%
  \BibitemOpen
  \bibfield  {author} {\bibinfo {author} {\bibfnamefont {D.}~\bibnamefont
  {Yanagishima}}\ and\ \bibinfo {author} {\bibfnamefont {Y.}~\bibnamefont
  {Maeno}},\ }\href {\doibase 10.1143/JPSJ.70.2880} {\bibfield  {journal}
  {\bibinfo  {journal} {Journal of the Physical Society of Japan}\ }\textbf
  {\bibinfo {volume} {70}},\ \bibinfo {pages} {2880} (\bibinfo {year}
  {2001})}\BibitemShut {NoStop}%
\bibitem [{\citenamefont {{Tafti}}\ \emph {et~al.}(2011)\citenamefont
  {{Tafti}}, \citenamefont {{Ishikawa}}, \citenamefont {{McCollam}},
  \citenamefont {{Nakatsuji}},\ and\ \citenamefont
  {{Julian}}}]{EuIridateExperiments}%
  \BibitemOpen
  \bibfield  {author} {\bibinfo {author} {\bibfnamefont {F.~F.}\ \bibnamefont
  {{Tafti}}}, \bibinfo {author} {\bibfnamefont {J.~J.}\ \bibnamefont
  {{Ishikawa}}}, \bibinfo {author} {\bibfnamefont {A.}~\bibnamefont
  {{McCollam}}}, \bibinfo {author} {\bibfnamefont {S.}~\bibnamefont
  {{Nakatsuji}}}, \ and\ \bibinfo {author} {\bibfnamefont {S.~R.}\ \bibnamefont
  {{Julian}}},\ }\href@noop {} {\bibfield  {journal} {\bibinfo  {journal}
  {ArXiv e-prints}\ } (\bibinfo {year} {2011})},\ \Eprint
  {http://arxiv.org/abs/1107.2544} {arXiv:1107.2544 [cond-mat.str-el]}
  \BibitemShut {NoStop}%
\bibitem [{\citenamefont {Nielsen}\ and\ \citenamefont
  {Ninomiya}(1981)}]{NielsenNinomiya}%
  \BibitemOpen
  \bibfield  {author} {\bibinfo {author} {\bibfnamefont {H.}~\bibnamefont
  {Nielsen}}\ and\ \bibinfo {author} {\bibfnamefont {M.}~\bibnamefont
  {Ninomiya}},\ }\href {\doibase 10.1016/0550-3213(81)90361-8} {\bibfield
  {journal} {\bibinfo  {journal} {Nuclear Physics B}\ }\textbf {\bibinfo
  {volume} {185}},\ \bibinfo {pages} {20 } (\bibinfo {year}
  {1981})}\BibitemShut {NoStop}%
\bibitem [{\citenamefont {Fritz}\ \emph {et~al.}(2008)\citenamefont {Fritz},
  \citenamefont {Schmalian}, \citenamefont {M\"uller},\ and\ \citenamefont
  {Sachdev}}]{FritzGrapheneConductivity}%
  \BibitemOpen
  \bibfield  {author} {\bibinfo {author} {\bibfnamefont {L.}~\bibnamefont
  {Fritz}}, \bibinfo {author} {\bibfnamefont {J.}~\bibnamefont {Schmalian}},
  \bibinfo {author} {\bibfnamefont {M.}~\bibnamefont {M\"uller}}, \ and\
  \bibinfo {author} {\bibfnamefont {S.}~\bibnamefont {Sachdev}},\ }\href
  {\doibase 10.1103/PhysRevB.78.085416} {\bibfield  {journal} {\bibinfo
  {journal} {Phys. Rev. B}\ }\textbf {\bibinfo {volume} {78}},\ \bibinfo
  {pages} {085416} (\bibinfo {year} {2008})}\BibitemShut {NoStop}%
\bibitem [{\citenamefont {{Arnold}}\ \emph {et~al.}(2003)\citenamefont
  {{Arnold}}, \citenamefont {{Moore}},\ and\ \citenamefont
  {{Yaffe}}}]{LeadingLogTransport}%
  \BibitemOpen
  \bibfield  {author} {\bibinfo {author} {\bibfnamefont {P.~B.}\ \bibnamefont
  {{Arnold}}}, \bibinfo {author} {\bibfnamefont {G.~D.}\ \bibnamefont
  {{Moore}}}, \ and\ \bibinfo {author} {\bibfnamefont {L.~G.}\ \bibnamefont
  {{Yaffe}}},\ }\href {\doibase 10.1088/1126-6708/2003/01/030} {\bibfield
  {journal} {\bibinfo  {journal} {Journal of High Energy Physics}\ }\textbf
  {\bibinfo {volume} {1}},\ \bibinfo {pages} {30} (\bibinfo {year} {2003})},\
  \Eprint {http://arxiv.org/abs/arXiv:hep-ph/0209353} {arXiv:hep-ph/0209353}
  \BibitemShut {NoStop}%
\bibitem [{\citenamefont {{Goswami}}\ and\ \citenamefont
  {{Chakravarty}}(2011)}]{GoswamiDiracTransport}%
  \BibitemOpen
  \bibfield  {author} {\bibinfo {author} {\bibfnamefont {P.}~\bibnamefont
  {{Goswami}}}\ and\ \bibinfo {author} {\bibfnamefont {S.}~\bibnamefont
  {{Chakravarty}}},\ }\href@noop {} {\bibfield  {journal} {\bibinfo  {journal}
  {ArXiv e-prints}\ } (\bibinfo {year} {2011})},\ \Eprint
  {http://arxiv.org/abs/1101.2210} {arXiv:1101.2210 [cond-mat.dis-nn]}
  \BibitemShut {NoStop}%
\bibitem [{\citenamefont {Kashuba}(2008)}]{KashubaGraphene}%
  \BibitemOpen
  \bibfield  {author} {\bibinfo {author} {\bibfnamefont {A.~B.}\ \bibnamefont
  {Kashuba}},\ }\href {\doibase 10.1103/PhysRevB.78.085415} {\bibfield
  {journal} {\bibinfo  {journal} {Phys. Rev. B}\ }\textbf {\bibinfo {volume}
  {78}},\ \bibinfo {pages} {085415} (\bibinfo {year} {2008})}\BibitemShut
  {NoStop}%
\bibitem [{Note1()}]{Note1}%
  \BibitemOpen
  \bibinfo {note} {We have used the fact that at particle-hole symmetry, the
  deviation from equilibrium is proportional to the sign of $\lambda $, and
  $f^0_\lambda (k) = 1- f^0_{-\lambda }(k)$.}\BibitemShut {Stop}%
\bibitem [{Note2()}]{Note2}%
  \BibitemOpen
  \bibinfo {note} {As mentioned in the introduction, $\alpha $ and $v_F$ take
  on their renormalized values at temperature $T$}\BibitemShut {NoStop}%
\end{thebibliography}%

\end{document}